\documentclass{aastex63}

\newcommand{\cm}{cm$^{-1}$}

\usepackage[version=3]{mhchem}
\usepackage[]{physics}

\received{\today}
\revised{}
\accepted{}
\submitjournal{Research Notes of the AAS}

\shorttitle{Insight for diatomic molecules to constrain $\mu$}
\shortauthors{Syme and McKemmish}

\begin{document}

\title{Computational insight into diatomic molecules as probes to measure the variation of the proton-to-electron mass ratio}

\correspondingauthor{Laura K. McKemmish}
\email{l.mckemmish@unsw.edu.au}

\author[0000-0003-3813-0823]{Anna-Maree Syme}
\affiliation{School of Chemistry, University of New South Wales, \\
2052 Sydney, Australia}

\author[0000-0003-1039-2143]{Laura K. McKemmish}
\affiliation{School of Chemistry, University of New South Wales, \\
2052 Sydney, Australia}



\begin{abstract}

Astrophysical molecular spectroscopy is an important means of searching for new physics through probing the variation of the proton-to-electron mass ratio, $\mu$.  New molecular probes could provide tighter constraints on the variation of $\mu$ and better direction for theories of new physics. Here we summarise our previous paper \citep{19SyMoCu.CN} for astronomers, highlighting the importance of accurate estimates of peak molecular abundance and temperature as well as spectral resolution and sensitivity of telescopes in different regions of the electromagnetic spectrum. Whilst none of the 11 astrophysical diatomic molecules we investigated showed enhanced sensitive rovibronic transitions at observable intensities for astrophysical environments, we have gained a better understanding of the factors that contribute to high sensitivities. From our results, CN, CP, SiN and SiC have shown the most promise of all astrophysical diatomic molecules for further investigation, with further work currently being done on CN. 

\end{abstract}



\section*{Introduction}

Currently, only 5\% of the observable Universe is understood by the standard model and general relativity. The incompatibility of these two theories has limited our understanding of the remaining 95\% of the universe. Several theories have been proposed to bridge this gap \citep{11Uz.mp2me}, with some predicting a temporal variation of fundamental constants, which can be tested by high accuracy measurements of observables that depend on these constants. Astrophysically, the proton-to-electron mass ratio, $\mu$, is observed across cosmological time through spectroscopy \citep{14JaBeUb.mp2me}; where the currently the tightest constraint is  $\dv{\mu}{t}/\mu < 2 \times 10^{-17} \text{yr}^{-1}$, assuming a linear change in time \citep{15KaUbMe.mp2me}.

The variation of $\mu$ is described by; 
\begin{equation}
    \frac{\Delta \nu}{\nu} = \frac{\nu_{obs}- \nu_{ref}}{\nu_{ref}} = K \frac{\Delta \mu}{\mu},
    \label{eqn:delta_nu}
\end{equation}
where $\frac{\Delta \mu}{\mu}$ is the fractional change in $\mu$, $K$ is the sensitivity coefficient, $\frac{\Delta \nu}{\nu}$ is the fractional change in the transition frequencies, $\nu_{obs}$ and $\nu_{ref}$ are observed and reference transition frequencies respectively. Equation \ref{eqn:delta_nu} assumes a variation in other fundamental constants will not affect $\frac{\Delta \mu}{\mu}$, which is suitable for the work done here \citep{11BeBoFl.mp2me}. $K$ is a theoretical value based on the transitions dependence on $\mu$. We assume that all baryonic matter will vary in the same manner, such that variations in molecular mass and $\mu$ \citep{07De.mp2me} are synonymous. While pure rotational, vibrational, and electronic transitions can easily be shown to have a sensitivity of -1, -0.5, and 0 respectively, we are searching for transitions that have enhanced sensitivities ($|K|>5$) from mixing of states or unusual features.

Previously the molecules used to constrain the variation in $\mu$ are \ce{H2}, \ce{HD}, \ce{CO}, \ce{NH3}, and \ce{CH3OH} \citep{14JaBeUb.mp2me}. \ce{NH3} and \ce{CH3OH} are preferred for their observed transitions with enhanced sensitivities, whereas \ce{H2}, \ce{HD}, and \ce{CO} are considered for their high abundance in astrophysical environments \citep{19SyMoCu.CN}. 



\section*{Screening diatomics}

A key aspect of our research was to bring together knowledge from chemistry and astronomy to optimise potential molecular probes. The selection criteria we devised has factors from both disciplines; 
\begin{itemize}
    \item Molecular properties and sensitivity coefficients of transitions;
    \item Intensity of the transitions and observational constraints;
    \item Resolution and the spectral range of required telescopes;
    \item Abundance and distribution of molecules in various astrophysical environments.. 
\end{itemize}

Use of these criteria was carried out in our previous publication \citep{19SyMoCu.CN}, however, our analysis was limited by the poor availability of consistently tabulated peak molecular abundance and temperatures as well as the spectral resolution and sensitivity at different wavelengths of telescopes. Sensitivity coefficients of the transitions, as well as frequency and intensity, were calculated from spectroscopic models \citep{19SyMoCu.CN}. Information of spectral range of telescopes was fairly accessible, however we found the lack of clarity for the resolution and sensitivity to be a obstacle, especially how that applied to the resolution of the spectral transitions observed and transition intensity requirements. The abundance and distribution of diatomics was arduously consolidated, and the scattered data sources made it difficult to have high confident in these abundances. With the understanding that astrophysical environments will have different distributions of molecules, we provided a guide to molecular prevalence.

\begin{figure}[hb]
    \centering
    \includegraphics[width=\textwidth]{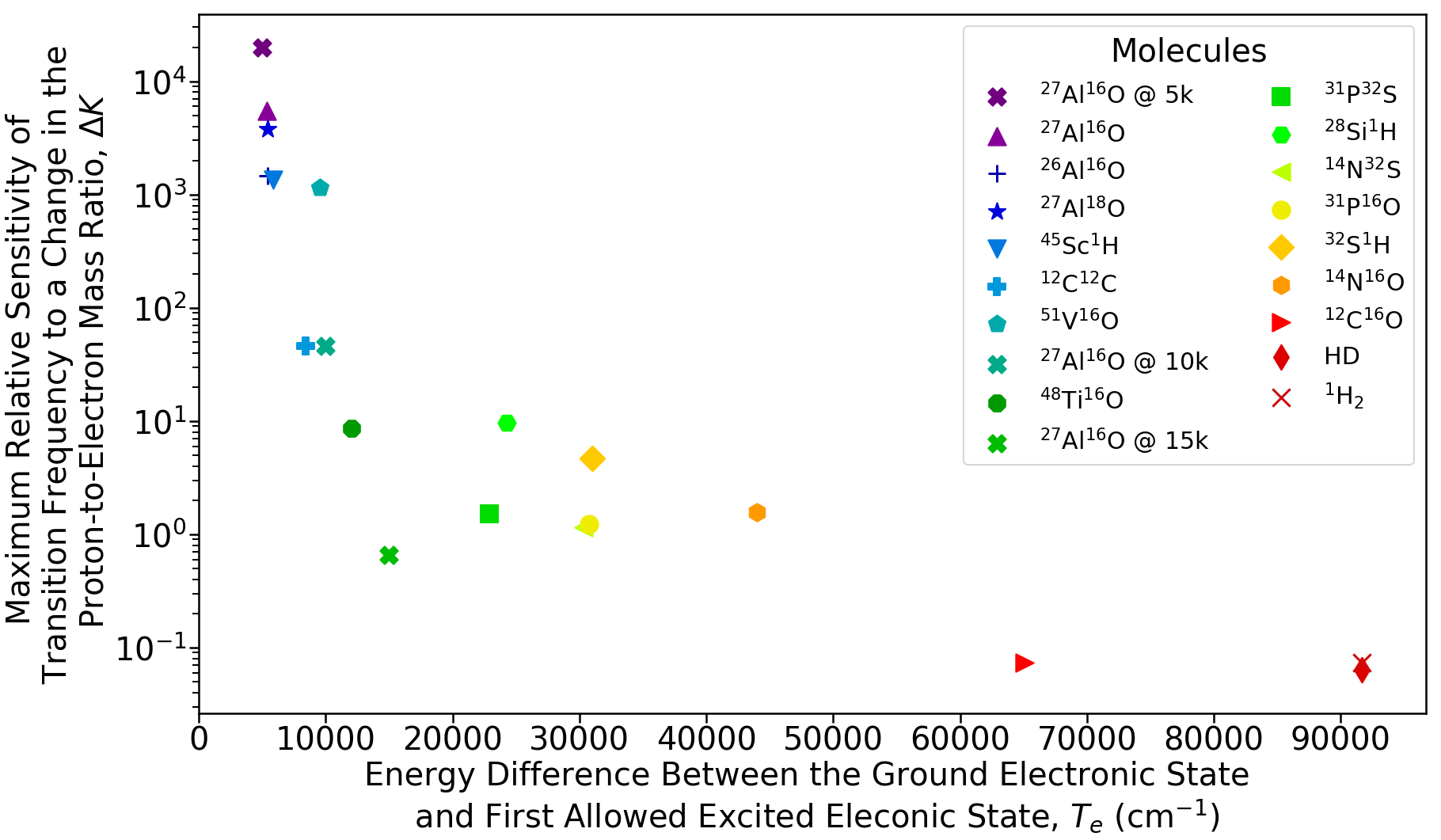}
    \caption{Relation between term energy, $T_e$ and the maximum $|\Delta K|$ for each diatomic examined, as well as the data points of \ce{^{27}Al^{16}O} where the spectroscopic model was altered by shifting the term energy, denoted with filled crosses. Also included is a combination of external data for \ce{CO}, \ce{HD}, and \ce{H2}. Figure adapted from \cite{19SyMoCu.CN}.}
    \label{fig}
\end{figure}


The 11 different diatomic molecules considered here are; \ce{^{12}C2}, \ce{^{32}S^{1}H}, \ce{^{14}N^{16}O}, \ce{^{14}N^{32}S}, \ce{^{31}P^{16}O}, \ce{^{31}P^{32}S}, \ce{^{28}Si^{1}H}, \ce{^{45}Sc^{1}H}, \ce{^{48}Ti^{16}O}, \ce{^{51}V^{16}O}, and \ce{^{27}Al^{16}O} including its isotopolgues \ce{^{26}Al^{16}O} and \ce{^{27}Al^{18}O}. Due to difficulties in observing astrophysical low-intensity transitions intensities less than $10^{-30}$ cm/molecule were omitted. Of the 11 diatomics considered here, none of them showed enhanced transitions at temperatures (10 K, 20 K, and 100 K) of astrophysical environments used for constraining the variation of $\mu$. When considering transitions within our intensity requirement at 1000 K we found \ce{SiH}, \ce{ScH}, \ce{C2}, \ce{TiO}, \ce{VO}, and all isotopologues of \ce{AlO} contained at least 1 transition with enhanced sensitivity. 


A minimum of two transitions are required to account for other astrophysical factors, therefore the relative sensitivity coefficients, $|\Delta K|$, must be maximised to constrain the variation of $\mu$ \citep{14JaBeUb.mp2me}. We observed the `term energy' $T_e$ (energy between the ground and excited electronic states) is consequential in the prevalence and extent of transitions with enhanced sensitivities by creating opportunities for more near-degenerate vibronic states. A correlation between $T_e$ and the maximum $|\Delta K|$ (figure \ref{fig}) for the diatomics with available data. The data for \ce{CO}, \ce{HD}, and \ce{H2} is possibly incomplete, with $T_e$ well-known, however the range of $K$ for these diatomics is limited by previously calculated values \citep{16UbKoEi.mp2me, 12SaNiBa.mp2me}. Figure \ref{fig} provides a guide to investigating other diatomics as potential probes for testing the variation of $\mu$. Lambda doubling, while shows promise is not explored, however diatomics that fit this criterion are also noted.

Curiously, none of our considered diatomics have significant abundance in molecular clouds, with 4 of them lacking any abundance estimate. Only NO and NS have been observed in extragalactic environments and neither showed enhanced sensitivities. All of the transitions that showed enhanced sensitivities were below 1000 \cm{} necessitating utilisation of infrared/radio telescopes and interferometers. 

The relationship between $T_e$ and maximum $|\Delta K|$ allows screening of all astrophysical diatomics and identify potential probes for further investigation. Of 43 known interstellar diatomics, we have identified;
\begin{itemize}
    \item 6 molecules (\ce{AlO}, \ce{SiC}, \ce{CP}, \ce{SiN}, \ce{C2}, \ce{CN}) that show promise for enhanced rovibronic transitions from low-frequency transitions between near degenerate vibronic states;
    \item 10 molecules (\ce{SO+}, \ce{PO}, \ce{OH}, \ce{CH}, \ce{SH}, \ce{NO}, \ce{NS}, \ce{HCl+}, \ce{SiC}, \ce{SiH}) with potential lambda doubling transitions;
    \item 2 molecules (\ce{SiS}, \ce{SiO}) with abundances similar to \ce{CO} which could be helpful as reference lines despite their predicted low sensitivity coefficients;
    \item 25 molecules  (\ce{HD}, \ce{HF}, \ce{NO+ }, \ce{N2}, \ce{PN}, \ce{CS}, \ce{SO}, \ce{AlCl}, \ce{O2}, \ce{NH}, \ce{OH+}, \ce{CH+}, \ce{CO+}, \ce{FeO}, \ce{TiO}, \ce{SiH}, \ce{AlF}, \ce{KCl}, \ce{CF+}, \ce{HCl}, \ce{NS+}, \ce{ArH+}, \ce{NaCl}, \ce{SH+}, \ce{CN$^-$}) with low potential for transitions with enhanced sensitivity, either lacking or low astronomical abundance.
\end{itemize}

The creation of a spectroscopic model for the extra-galactically observed \ce{CN} radical \citep{20SyMc.CN}, intent on discerning transitions with enhanced sensitivities and higher intensities at lower temperatures, is underway.

\bibliography{bib}{}
\bibliographystyle{aasjournal}



\end{document}